\documentstyle{mn}

\include{psfig}
\begin{document}

\title[Accretion disc outbursts]{Accretion disc outbursts: a new version of
an old model}

\author[J.-M. Hameury et al.]{Jean-Marie Hameury$^1$, Kristen Menou$^{2,3}$
Guillaume Dubus$^3$, Jean-Pierre Lasota$^3$ \cr
and Jean-Marc Hur\'e$^4$
\\$^1$ UMR 7550 du CNRS, Observatoire de Strasbourg,
	11 rue de l'Universit\'e, F-67000 Strasbourg, France
\\$^2$ Harvard-Smithsonian Center for Astrophysics, 60 Garden
	Street, Cambridge, MA 02138, USA
\\$^3$ UPR 176 du CNRS, D\'epartement d'Astrophysique Relativiste
	et de Cosmologie, Observatoire de Paris, Section de Meudon,
\\	F-92195 Meudon C\'edex, France
\\$^4$ URA 173 du CNRS, D\'epartement d'Astrophysique Extragalactique
	et de Cosmologie, Observatoire de Paris, Section de Meudon,
\\F-92195
	Meudon C\'edex, France}

\maketitle

\begin{abstract}
We have developed 1D time-dependent numerical models of accretion discs,
using an adaptive grid technique and an implicit numerical scheme, in which
the disc size is allowed to vary with time. The code fully resolves the
cooling and heating fronts propagating in the disc. We show that models in
which the radius of the outer edge of the disc is fixed produce incorrect
results, from which probably incorrect conclusions about 
the viscosity law have been inferred. 
In particular we show that outside-in outbursts are possible
when a standard bimodal behaviour of the Shakura-Sunyaev viscosity parameter
$\alpha$ is used. We also discuss to what extent insufficient grid
resolutions have limited the predictive power of previous models. We find
that the global properties (magnitudes, etc. ...) of transient discs can be
addressed by codes using a high, but reasonable, number of fixed grid points.
However, the study of the detailed physical properties of the transition
fronts generally requires resolutions which are out of reach of fixed grid
codes. It appears that most time-dependent models of accretion discs
published in the literature have been limited by resolution effects, improper
outer boundary conditions, or both.
\end{abstract}


\section{Introduction}

The thermal-viscous accretion-disc instability model is more than 15 years
old (see Cannizzo \shortcite{can93b} for a historical overview). It is widely
accepted that it provides the correct description of dwarf-nova outbursts and
of (`soft') X-ray transient events. When, however, observations of these
systems are compared with predictions of the model, the agreement is far from
perfect (e.g. Lasota \& Hameury \shortcite{lh98} and references therein). It
is sometimes also unclear what the predictions of the model are. One of the
reasons for these uncertainties is the existence of various, different,
versions of the model. From the very beginning these versions of the disc
instability model differed in assumptions about viscosity and boundary
conditions; they differed in the amount of matter accreted during the
outburst, the shapes of light-curves, etc. (see Cannizzo \shortcite{can93b}).
At that time these differences seemed to be less important than the
differences between the disc instability model and the competing,
mass-transfer instability model \cite{bp81}. The exponentially decaying tails
of theoretical light curves predicted by the mass-transfer model were thought
to contradict observations \cite{can93b} and the outer disc radius behaviour
during and after outbursts seemed to favour the disc instability model
\cite{io92}. The demise of the mass-transfer instability model was, however,
caused by the lack of a physical mechanism which would trigger it.

With one model left it became important to establish just what its
predictions are, and not merely whether it is better (or worse) than the
competing model \cite{pvw86}. The first systematic study of the disc
instability model was presented by Cannizzo \shortcite{can93a}, who analysed
the importance of various terms in the disc evolution equations and the
influence of the numerical grid resolution on the outburst properties.
Ludwig, Meyer-Hofmeister \& Ritter \shortcite{lmr94} studied general
properties of disc outbursts, such as the location of the instability that
triggers them. Recently Ludwig \& Meyer \shortcite{lm98} analysed
non-Keplerian effects which may arise during front propagation. The general
conclusions of this group of studies were that non-Keplerian effects are
negligible, that a few hundred grid points provide a sufficient resolution
for the calculation results to be independent of the number of grid points,
and, finally, that with the usual assumption (see Smak (1984b) of a jump in
the value of the viscosity parameter $\alpha$, the model produces only
`inside-out' outbursts, i.e. outbursts starting in the inner disc regions.
This last conclusion, if true, would entail changing the standard viscosity
law (in which the $\alpha$ parameter is constant in the hot and cool branch
of the $\Sigma$ - $T_{\rm eff}$ curve) because outburts starting in the outer
disc regions are clearly observed in classical dwarf-nova system SS Cyg
\cite{mau96}. This law already had to be modified when it was found
\cite{sma84b} that in order to get lightcurves similar to those observed in
dwarf novae, $\alpha$ in outburst had to be larger than $\alpha$ in
quiescence.

The absence of `outside-in' outbursts in these studies, however, is 
just the result of keeping the outer disc radius constant in the calculations
(Section 4.1) \cite{io92,sma84b}; from this point of view, there is no reason
to modify the viscosity prescription. This does not in itself prove that
changes in viscosity are correctly described by the bimodal behaviour of the
$\alpha$-parameter (see e.g. Gammie \& Menou 1998), but the reasons given for
preferring other versions (the exponential decay from outburst being the
principal one) are not compelling, and these versions involve more
fundamental changes in the disc physics (see Lasota \& Hameury 1998 for
discussion and references).

For example, Cannizzo, Chen \& Livio \shortcite{ccl95} use the formula
$\alpha = \alpha_0 \left(H/R \right)^n$, but to make the model work they have
to `switch off' convection. It is interesting, therefore, to recall that
Faulkner, Lin \& Papaloizou (1983) found dwarf nova outburst with $\alpha$
constant, but their model was criticized \cite{can93b} because they claimed
that convection has only a minor influence on the energy transport in the
disc. These are not formal problems because the constant $\alpha$ models
predict optically thin quiescent discs, whereas in bimodal $\alpha$ models
the quiescent disc is optically thick. There is observational evidence that
dwarf nova discs in quiescence are optically thin (see Horne, 1993 and
references therein).

Conclusions about the number of grid points required to get
resolution-invariant results seem, on inspection, too optimistic, especially
because fronts are not resolved, a point which is particularily worrying for
the heating fronts.

The present situation of the disc instability model seems to be confused.
Various versions are based on different assumptions about the physical
processes in the disc and numerical codes suffer either from incorrect
boundary conditions or from insufficient resolution or from both. Quite
often, in the case of explicit codes the resolution is limited by the
required computer time.

In this article we describe a numerical model of time-dependent accretion
discs, using an adaptive grid technique and an implicit numerical scheme, in
which the disc size is allowed to vary with time. This numerical scheme
allows rapid calculations of disc outburst cycles at very high resolution.
These properties allow an easy comparison with other versions of the model
and a systematic study of its various assumptions.

In the near future we will use our code to model various properties of dwarf
novae and X-ray transients. The model was alread used to model properties and
outbursts of the dwarf-nova WZ Sge \cite{lhh,hlh} and the rise to outburst of
the X-ray transient GRO J1655-40 \cite{hlmn97}.

In \S 2 we discuss the time-dependent equations describing the disc radial
structure and the implicit method used to solve them with a high spatial
resolution. The vertical structure of the disc, and hence the heating and
cooling terms that enter the time-dependent energy equation, are considered
in \S 3. In \S 4 we present the results of our calculations and we discuss
the importance of having sufficient numerical resolution and a correct
boundary condition at the outer edge of the disc.

\section{Time-dependent accretion discs}

\subsection{Disc equations}

The basic equations for mass and angular momentum conservation in a
geometrically thin accretion disc can be written as:
\begin{equation}
{\partial \Sigma \over \partial t} = - {1 \over r} {\partial \over \partial
r} (r \Sigma v_{\rm r}) + {1 \over 2 \pi r} {\partial \dot{M}_{\rm ext} \over
\partial r}
\label{eq:consm}
\end{equation}
and
\begin{eqnarray}
j{\partial \Sigma \over \partial t} = - {1 \over r} {\partial \over \partial
r} (r \Sigma j v_{\rm r}) & + & {1 \over r} {\partial \over \partial r}
\left(- {3 \over 2} r^2 \Sigma \nu \Omega_{\rm K} \right) + \nonumber \\
& & { j_{\rm k} \over 2 \pi r} {\partial \dot{M}_{\rm ext} \over \partial r} -
{1 \over 2 \pi r} T_{\rm tid}(r),
\label{eq:consj}
\end{eqnarray}
where $\Sigma$ is the surface column density, $\dot{M}_{\rm ext}(r)$ is the
rate at which mass is incorporated into the disc at point $r$, $v_{\rm
r}$ the radial velocity in the disc, $j = (GM_1r)^{1/2}$ is the specific
angular momentum of material at radius $r$ in the disc,
$\Omega_K=(GM_1/r^3)^{1/2}$ is the Keplerian angular velocity ($M_1$ being
the mass of the accreting object), $\nu$ is the kinematic viscosity
coefficient, and $j_{\rm k}$ the specific angular momentum of the material
transfered from the secondary. $T_{\rm tid}$ is the torque due to tidal
forces, for which we use the value of Smak \shortcite{sma84b}, derived
from the determination of tidal torques by Papaloizou \& Pringle
\shortcite{pp77}):
\begin{equation}
T_{\rm tid} = c \omega r \nu \Sigma \left( {r \over a} \right)^{5},
\label{eq:defT}
\end{equation}
where $\omega$ is the angular velocity of the binary orbital motion, $c$ is a
numerical coefficient taken so as to give a stationary (or time
averaged) disc radius equal to some chosen value, and $a$ is the binary
orbital separation.

Equation~(\ref{eq:consj}) is the standard form of angular momentum
conservation, in which it is assumed that the azimuthal velocity $v_\phi$ has
the Keplerian value $v_{\rm K}$. Ludwig \& Meyer \shortcite{lm98} have
considered deviations from Keplerian motion arising from strong pressure
gradients (i.e. in the heating and cooling fronts) and found that these
deviations only have a marginal influence on the outburst behaviour.

The energy conservation equation is taken as:
\begin{equation}
{\partial T_{\rm c} \over \partial t} = { 2 (Q^ + -Q^- + J) \over C_P \Sigma}
 - {\Re T_{\rm c}
\over \mu C_P} {1 \over r} {\partial (r v_{\rm r}) \over \partial r} -
v_{\rm r} {\partial T_{\rm c} \over \partial r},
\label{eq:heat}
\end{equation}
where $Q^+$ and $Q^-$ are the surface heating and cooling rates respectively.
They are usually taken as $Q^+=(9/8) \nu \Sigma \Omega_{\rm K}^2$ and $Q^- =
\sigma T_{\rm eff}^4$, $T_{\rm eff}$ being the effective temperature (e.g.
Cannizzo \shortcite{can93a}). As described in \S 3 we use slightly different
forms of these rates. The term $J$ accounts for the radial energy flux
carried either by viscous processes or by radiation. In the following, we
neglect the radiative flux which, according to Ludwig \& Meyer
\shortcite{lm98}, is negligible, while according to Cannizzo
\shortcite{can93a} it is comparable to the viscous radial flux.

If viscosity is due to turbulence, $J$ can be estimated in the framework of
the $\alpha$ parametrization. The flux carried in eddies with characteristic
velocity $v_{\rm e}$ and size $l_{\rm e}$, is:
\begin{equation}
F_{\rm e} = C_P \Sigma v_{\rm e} {\partial T_{\rm c} \over \partial r} l_{\rm
e} = {3 \over 2} \nu C_P \Sigma {\partial T_{\rm c} \over \partial r},
\label{eq:fturb}
\end{equation}
in which case
\begin{equation}
J = 1/r \partial / \partial r (r F_{\rm e}).
\end{equation}
A similar expression is obtained for radiative transport. Note that this
expression differs slightly from that used by Cannizzo \shortcite{can93a}
which cannot be written as the divergence of a flux. Other prescriptions for
$J$ exist, in time-dependent simulations; they give results very
similar to those obtained using Eq.~(\ref{eq:fturb}).

\subsection{Boundary conditions}

So far there is no clear understanding of the physics of the interaction
between the disc and the stream (see e.g. Armitage \& Livio, 1997),
and a fraction of the stream may spill over the disc.
A proper treatment of the precise way in which matter is incorporated into
the disc is far beyond our scope, and we use here the simplest, but very
reasonable assumption that mass addition at the outer edge of the disc occurs
in a very narrow region, so that the disc edge is very sharply defined. We
can then write $\dot{M}_{\rm ext}(r) = \dot{M}_{\rm tr} \delta (r_0(t)-r)$
and $\Sigma = \Sigma_0 Y(r_0(t)-r)$, where $\dot{M}_{\rm tr}$ is the mass
transfer rate from the secondary star, $Y$ is the Heavyside function,
$\delta$ is the Dirac function and $\Sigma_0$, the surface column density, is
a smoothly varying function. The cancelation of the $\delta$ terms in the
equation for mass and angular momentum conservation yields two boundary
conditions, which can be written in the form:
\begin{equation}
\dot{M}_{\rm tr} = 2 \pi r \Sigma_0 (\dot{r}_0 - v_{\rm r,0})
\label{eq:bc1}
\end{equation}
and
\begin{equation}
\dot{M}_{\rm tr} \left[ 1 - \left( {r_{\rm k} \over r_0}\right)^{1/2}
\right] = 3 \pi \nu \Sigma_0,
\label{eq:bc2}
\end{equation}
where the index 0 denotes quantities measured at the outer edge, and $r_{\rm
k}$ is the circularization radius, i.e. the radius at which the Keplerian
angular momentum is that of the matter lost by the secondary star, and
$\dot{r}_0$ is the time derivative of the outer disc radius. It is worth
noting that the presence of a torque $T_{\rm tid}$ is required in this
formulation, and it can be easily seen that no steady solutions exist when
$T_{\rm tid}=0$.

Conditions (\ref{eq:bc1}, \ref{eq:bc2}) take into account the fact that the
outer edge of the disc can vary with time, and its position is controlled by
the the tidal torque $T_{\rm tid}$. Variations of the disc radius are
observed during outburst cycles in dwarf novae
\cite{sma84a,od86,wmr89,wmh93}, with a rapid rise during the outburst and a
decrease of the order of 20 percent during decline; such variations have been
considered as a strong argument in favour of the disc instability model for
dwarf novae, as opposed to the mass transfer instability model. The presence
of `superhumps' during SU UMa's superoutbursts is explained by the disc
radius becoming larger than the radius for the 3:1 resonance (see e.g Frank,
King \& Raine \shortcite{fkr92}). Osaki \shortcite{o96} assumes that the
superoutbursts themselves are due to a `tidal-thermal instability' which sets
in when the disc's outer radius reaches the 3:1 resonance zone; see however
Smak \shortcite{sma96}.

Nevertheless it is often assumed, when modeling the normal outbursts, that
the outer edge of the disc is fixed at a given radius, in which case
Eq.~(\ref{eq:bc1}) is used with $\dot{r}_0 = 0$, and Eq.~(\ref{eq:bc2}) is
replaced by $r = r_0$; the tidal torque $T_{\rm tid}$ is also neglected in
Eq.~(\ref{eq:consj}). This is equivalent to assuming that $T_{\rm tid}$ is
negligible at $r < r_0$ and becomes infinite at $r = r_0$. In view of the
steep functional dependence of $T_{\rm tid}(r)$, this might seem a reasonable
approximation; however, we shall show that this is not the case at all, and
that results obtained with both boundary conditions 
differ very significantly. An
intermediate formulation has been proposed by Mineshige \& Osaki
\shortcite{mo85}, who assume that the viscous stresses vanish at a fixed
outer radius. This enables matter carrying angular momentum to leave the disc
at its outer edge, at a rate which is comparable to the mass transfer rate.

We take, as usual, $\Sigma = 0$ at the inner edge of the disc (more exactly,
$\Sigma$ equal to a small value).

The thermal equation being a second order partial differential equation in
$r$, two boundary conditions are required. However, except across a
transition front between a hot and a cool region, the dominant terms in
Eq.~(\ref{eq:heat}) are $Q^+$ and $Q^-$. The highest order terms in the
thermal equation are therefore negligible in almost all of the disc; the
solutions of such equations are known to develop boundary layers, which adapt
the internal solution (in which the highest order terms are neglected) to the
boundary conditions. These boundary conditions are thus of no physical
importance, and should be such as to minimize numerical difficulties. We have
here taken $\partial T_{\rm c} / \partial r = 0$ at both edges of the disc.

\subsection{Numerical method}

We solve the set of equations (\ref{eq:consm}, \ref{eq:consj}, \ref{eq:heat})
using a method described by Eggleton \shortcite{e71}. This method, which uses
a variable mesh size, is a generalization of the Henyey method \cite{h59},
designed to solve the set of non-linear equations describing the internal
structure of stars. The solutions of these equations have steep gradients,
both at their surface and in the vicinity of the thin burning shells which
appear after their evolution off the main sequence; moreover, the position of
the shells is not fixed, but may vary relatively rapidly across the stellar
envelope.

The natural abscissa, $r$ in our case, is considered as an unknown variable
of a new parameter $q$ in the range 0 -- 1 over the grid. The variable $r$
varies according to:
\begin{equation}
{dr \over dq } = \Phi \times W(r, T_{\rm c}, \Sigma, ...), \label{eq:griddef}
\end{equation}

where W is a function which becomes small when the radial derivatives of
$T_{\rm c}$ or $\Sigma$ become large, and $\Phi$ is a normalization constant
which adjusts the grid to the physical range covered by the variable $r$, so
that
\begin{equation}
{d\Phi \over dq }= 0.
\label{eq:phiconst}
\end{equation}
Here, we take
\begin{equation}
W^{-2} = 1 + 0.05 \left[{\partial \ln (\nu \Sigma r^{1/2}) \over \partial \ln
r}\right]^2 + 0.1 \left[ {\partial \ln T_{\rm c} \over
\partial \ln r}\right]^2.
\end{equation}
The thermal and viscous equations (\ref{eq:consm}, \ref{eq:consj},
\ref{eq:heat}) are equivalent to a set of four first order differential
equations. These equations plus the two equations (\ref{eq:griddef},
\ref{eq:phiconst}) defining the grid are discretized and solved using a
generalized Newton method. They can be written in the form:
\begin{equation} {d F_i(f^j) \over dq} = G_i(f^j), \qquad (i=1,l),
\end{equation}
where $l=6$ is the number of equations, the $f^j$ are the 6 variables
$r$, $\Phi$, $T_{\rm c}$, $\Sigma$, $\partial (\nu \Sigma r^{1/2}) / \partial
r$ and $\partial T_{\rm c}/\partial r$, and the $G_i$ terms contain the
time-derivatives. The discretized equation takes the form:
\begin{equation}
F_i(f_k^j) - F_i(f_{k-1}^j) - \delta q [\beta_i G(f_k^j) +(1-\beta_i)
G(f_{k-1}^j)] = 0,
\label{eq:discr}
\end{equation}
where $\delta q = 1/(N-1)$ is the mesh size, $f_k^j$ is the value of the
quantity $f^j$ at the grid point $k$ ($k=1,N$), and the $\beta_i$ are weights
in the range $[0,1]$. Values of $\beta_i$ equal to 0.5 lead to a second
order scheme. The set of equations (\ref{eq:discr}) is linearized (including
Eq.~(\ref{eq:griddef}-\ref{eq:phiconst}) by numerical differentiation. This
method has the advantage of being fully implicit, but convergence may be a
problem when the initial guess is not close enough to the actual solution.

Whereas in Eggleton \shortcite{e71} the time derivatives are expressed in
their Eulerian form, we directly estimate the Lagrangean values by using
either a linear or a cubic spline interpolation of the quantities calculated
at the previous timestep, on the previous grid. The latter is more accurate,
but requires a higher number of points when a transition front is present in
the disc in order to avoid numerical oscillations.

This formulation does not guarantee that the disc mass is a conserved
quantity within roundoff errors. Instead, the disc mass has to be explicitly
calculated by integrating $\Sigma$ over the disc extension, and deviations
from mass conservation (easy to calculate given the inner and outer mass
accretion rates in the disc, together with the time step of integration)
provide valuable information on the quality of the code. We checked that all
adaptive grid models presented here do indeed conserve mass accurately,
except when we use too small a number of grid points (typically 100 or less).
In the latter case, convergence is often a problem.

In order to improve the stability of the numerical scheme, we have used
values of $\beta_i$ equal to 0 for the two equations defining the derivatives
of $T_{\rm c}$ and $\Sigma$, and to 1 for the two equations defining their
second derivatives when the radius of the outer edge of the disc is kept
fixed; we have also calculated the Lagrangean derivatives using linear
interpolations. However, when $r_0$ is allowed to vary, mass conservation is
more difficult to enforce; we have taken all $\beta_i$ equal to 0.5 in that
case, and used cubic splines to calculate the time derivatives. The timestep
is chosen on empirical grounds and depends on the quality of the previous
convergence. We find that initiating the disc with a hot steady solution or a
globally cold state with an arbitrary $\Sigma$ profile gives, after a short
transient period, equivalent results.

\section{Determination of the effective temperature}

The vertical structure equations are very similar to those describing the
internal structure of a star, with the notable exception that energy is
dissipated everywhere in the vertical structure, including the optically thin
regions, and that the vertical gravity comes from the central object.

In order to follow the time evolution of accretion discs, one needs to know
the thermal imbalance ($Q^+ - Q^-$) as a function of the central temperature
$T_{\rm c}$ and the surface column density $\Sigma$ at any radius in the
disc. One should then be able to solve the full 2D energy transfer problem,
which, in the thin disc approximation, is assumed to decouple into radial
(Eq.~(\ref{eq:heat})) and vertical equations. The latter, together with the
reasonable assumption of vertical hydrostatic equilibrium, are as follows
\cite{sma84b}:
\begin{eqnarray}
\lefteqn{{dP \over dz} = -\rho g_{\rm z} = -\rho \Omega_{\rm K}^2 z, }
\label{eq:stra}\\
\lefteqn{{d \varsigma \over dz} = 2 \rho,} \label{eq:strd} \\
\lefteqn{{d\ln T \over d  \ln P} = \nabla,} \label{eq:strb}\\
\lefteqn{{dF_{\rm z} \over dz } = {3 \over 2} \alpha \Omega_{\rm K} P +
{dF_t \over dz},} \label{eq:strc}
\end{eqnarray}
where $P$, $\rho$ and $T$ are the pressure, density and temperature
respectively, $\varsigma$ is the surface column density between $-z$ and
$+z$, $g_{\rm z} = \Omega_{\rm K}^2 z$ the vertical component of gravity,
$F_{\rm z}$ the vertical energy flux and $\nabla$ the temperature gradient of
the structure. This is generally radiative, with $\nabla=\nabla_{\rm rad}$,
given by:
\begin{equation}
\nabla_{\rm rad} = {\kappa P F_{\rm z} \over 4 P_{\rm rad} c g_{\rm z}},
\end{equation}
$P_{\rm rad}$ being the radiative pressure. When the radiative
gradient is superadiabatic, $\nabla$ is convective
($\nabla=\nabla_{\rm conv}$). The convective gradient is calculated in the
mixing length approximation, with a mixing length taken as $H_{\rm ml} =
\alpha_{\rm ml} H_P$, where $H_P$ is the pressure scale height:
\begin{equation}
H_P = {P \over \rho g_{\rm z} +(P \rho)^{1/2} \Omega_{\rm K} },
\end{equation}
which ensures that $H_P$ is smaller than the vertical scale height of the
disc. We have \cite{p69}:
\begin{equation}
\nabla_{\rm conv} = \nabla_{\rm ad} + (\nabla_{\rm rad} - \nabla_{\rm ad}) Y
(Y+A)
\end{equation}
where $\nabla_{\rm ad}$ is the adiabatic gradient, and $Y$ is the solution of
the cubic equation:
\begin{equation}
{9 \over 4} {\tau_{\rm ml}^2 \over 3 +\tau_{\rm ml}^2} Y^3 + VY^2 + V^2 Y -V
= 0
\end{equation}
where $\tau_{\rm ml} = \kappa \rho H_{\rm ml}$ is the optical depth of
the convective eddies. The coefficient $V$ is given by:
\begin{eqnarray}
V^{-2} = \left( {3 + \tau_{\rm ml}^2 \over 3 \tau_{\rm ml}}\right)^2 {g_{\rm
z}^2 H_{\rm ml}^2 \rho^2 C_P^2 \over 512 \sigma^2 T^6 H_P} \left( {\partial
\ln \rho \over \partial \ln T} \right)_P \nonumber \\
\times (\nabla_{\rm rad} - \nabla_{\rm ad})
\end{eqnarray}
Here we take $\alpha_{\rm ml} = 1.5$, which is appropriate for solar type
stars ($\alpha_{\rm ml}$ ranges from 1 to 2 in solar models, see e.g. Guzik \&
Lebreton \shortcite{gl91} and Demarque and Guenther \shortcite{dg91}). One
should note that $\alpha_{ml} = 1$ is generally used in the literature; as
the disc vertical structure depends on the assumed $\alpha_{\rm ml}$
\cite{can93b}, we expect some
difference from previous
results. $F_t$ is a time-dependent contribution that includes terms
resulting from heating/cooling and contraction/expansion. This contribution
is basically unknown, but it is most often assumed as proportional to $P$; if
this is the case, then Eq.~(\ref{eq:strc}) can be replaced by:
\begin{equation}
{dF_{\rm z} \over dz } = {3 \over 2} \alpha_{\rm eff} \Omega_{\rm K}
P,
\label{eq:strcc}
\end{equation}
where $\alpha_{\rm eff}$ may be considered as some effective viscosity. This
coefficient is not known {\it a priori} and is generally different from the
true viscosity parameter; the difference $\alpha - \alpha_{\rm eff}$ measures
the departure from thermal equilibrium. The vertical structure, and hence the
disc effective temperature at any given point, can therefore be obtained
assuming that the disc is in thermal equilibrium, but with some unspecified
$\alpha_{\rm eff}$ different from $\alpha$.

Even though this approximation is not unreasonable, and can be shown to be
valid in the simplest case, that of a perfect gas and homologous cooling and
contraction \cite{sma84b}, it is by no means justified and constitutes a
significant drawback to the model. Other approximations have been used in the
past; it is for example possible to assume that $F_{\rm z}$ has a given
vertical profile; this is the approach of Mineshige \& Osaki \shortcite{mo83}
who assumed that $F_{\rm z}$ increases linearly with altitude up to a
saturation value. This is not drastically different from assuming that
$F_{\rm t}$ is proportional to pressure, since in both cases, energy is
released in the densest regions of the disc, i.e. within one or a few scale
heights from the mid plane. We have compared the results given in both cases,
and we found that, for a given $(\Sigma,T_{\rm c}$), $T_{\rm c}$ being the
central temperature, the effective temperatures obtained agreed to within 5
to 10 \%. Therefore, the cooling fluxes in time-dependent accretion discs
are not determined better than to within 50\%.

It would be highly desirable to have a better determination of $Q^-$;
however, this would require at least the introduction of explicitly
time-dependent
terms in the vertical structure equations, and that would couple the radial
and vertical structure of the disc. This would enormously increase the
required amount of computing time, making it very difficult to follow a
complete outburst cycle with a reasonable number of radial grid points.

The viscous dissipation term $3/2 \alpha \Omega_K P$ in Eq.~(\ref{eq:strc})
is a direct consequence of the assumption that the radial-azimuthal component
of the viscous stress tensor can be written $\tau_{r\phi}=-\alpha P$
\cite{ss73}. One should also note that the $\alpha$
prescription is used here in its local version, i.e. the local energy
dissipation is assumed to be proportional to the pressure, which is by
no means obvious.

Our approach is to determine $Q^-$ for a number of values of $T_{\rm c}$,
$\Sigma$, and $r$ and then perform a linear interpolation. In what follows,
we have taken 70 equally logarithmically spaced values for $r$ (190, 120 for
$T_{\rm c}$, $\Sigma$ respectively). This may seem time-consuming, but once
the initial grid has been calculated, it is much more precise than the
approach in which a sparse grid is used to construct analytic fits of $Q^-$.
The $\Sigma-T_{\rm eff}$ ``S-curves'' are found numerically by looking in the
grid for values of $\Sigma$, $T_{\rm c}$, $\alpha$ and $r$ which satisfy the
thermal equilibrium condition.

In thermal equilibrium the heating term $Q^+$ is given by:
\begin{equation}
Q^+ = {3 \over 2} \alpha \Omega_K \int_0^{+\infty} P dz, \label{eq:qplus}
\end{equation}
which is close, but not exactly equal, to the heating term $(9/8) \nu
\Sigma \Omega_{\rm K}^2$ generally used in the literature. Setting $\nu = 2/3
\alpha c_{\rm s}^2 / \Omega_{\rm K}$, and assuming that $c_{\rm s}$ is
constant, one recovers the classical expression. This slight difference might
appear as a minor inconsistency, but it may lead to unwanted numerical
effects, since the equilibrium situation found when solving the radial
structure would be different from that found in this section, and the disc
may become unstable for a surface density slightly different from the
expected critical value. For this reason, we prefer to use the exact 
value of the integral in (\ref{eq:qplus}) rather than its approximation; it
is stored and interpolated in order to provide an accurate $Q^+$ to the
time-dependent runs.

We solved the vertical structure equations in two different approximations
for the radiative transfer of energy, namely in the optically thick
approximation and in the grey atmosphere approximation.

\subsection{Optically thick case}

The set of equations (\ref{eq:stra}-\ref{eq:strb}, \ref{eq:strcc}) is
integrated between the disc midplane and the photosphere, defined as
the point where:
\begin{equation}
\kappa_{\rm R} P = {2 \over 3} g_{\rm z}, \label{eq:phote}
\end{equation}
where $\kappa_{\rm R}$ is the Rosseland mean opacity.

The other boundary conditions are $z = 0$, $F_{\rm z} = 0$, $T = T_{\rm c}$,
$\varsigma = 0$ at the disc midplane, and $F_{\rm z} = \sigma T^4$,
$\varsigma = \Sigma$ at the photosphere.

\subsection{Grey atmosphere approximation}

For low values of the surface column density, the disc optical depth is no
longer large. An accurate solution of the vertical disc structure would
require an excessive amount of computing time, since frequency-dependent
opacities would have to be taken into account for consistency. Instead, we
chose to integrate the set of equations (\ref{eq:stra}--\ref{eq:strd}) down
to optically thin regions, but we use a grey atmosphere approximation in
which the temperature varies as $T^4 = T_{\rm s}^4 (1/2 + 3/4 \tau)$, where
$T_{\rm s}$ is the surface temperature and $\tau$ the optical depth, given by:
\begin{equation}
{d\tau \over dz} = \kappa \rho.
\label{eq:deftau}
\end{equation}
For optical depths larger than unity, one should use Rosseland opacities
$\kappa_{\rm R}$, whereas at small optical depths, the Planck mean opacities
$\kappa_{\rm P}$ are relevant. Here we take
\begin{equation}
\kappa = {\tau_{\rm e}^2 \over 1 + \tau_{\rm e}^2} \kappa_{\rm R} + {1 \over
1 + \tau_{\rm e}^2} \kappa_{\rm P},
\end{equation}
where $\tau_{\rm e} = \Sigma \kappa_{\rm R}$ is the estimated disc optical
depth. This $T(\tau)$ relation leads to the following boundary condition at
the surface:
\begin{equation}
F_{\rm z} = 2 \sigma \int_0^{\tau_{\rm s}} T(\tau)^4 [E_2(\tau) + E_2(2
\tau_{\rm s} -\tau)] d \tau,
\end{equation}
where $E_2$ is the exponential-integral function. This relation can be
approximated to better than $2 \%$ by
\begin{equation}
F_{\rm z} = 2 \sigma T_{\rm s}^4 (1 - e^{-2 \tau_{\rm s}} -0.84 \tau_{\rm
s}^{3/2} e^{-2 \tau_{\rm s}}), \label{eq:photm}
\end{equation}
where $\tau_{\rm s}$ is the optical depth from the midplane to the surface.
The grey atmosphere approximation is rather crude (see e.g. Shaviv \& Wehrse
1991; Hubeny 1990); moreover, it does not account for energy deposition in
optically thin layers. However, calculations of the vertical disc structure
using the Shaviv \& Wehrse (1991) radiative transfer code show that the grey
approximation is quite good, especially in view of uncertainties in $Q^-$
mentioned above \cite{ihls98}. In any case, the grey atmosphere
approximation is required if one is willing to calculate a large number of
such disc structures in a reasonable amount of computing time.

The two photospheric boundary conditions (\ref{eq:phote}) and $F_{\rm z} =
\sigma T^4$ are replaced by (\ref{eq:photm}) and $\rho$ vanishingly small (we
take here $\rho = 10^{-10}$ g cm$^{-3}$). The optical depth has to be
explicitly integrated in the vertical structure. Consequently,
Eq.~(\ref{eq:deftau}) is added to the set of equations that we solve, with
the associated boundary condition $\tau = 0$ at the disc surface.

\subsection{Numerical method}

We solve the set of equations (\ref{eq:stra}--\ref{eq:strc}) for given values
of the surface density $\Sigma$ and central temperature $T_{\rm c}$ using the
same adaptive grid numerical method as for the disc equations. The grid is
now defined by:
\begin{equation}
W^{-2} = {4 \over z_{\rm norm}^2} + \left({1 \over \Sigma} {d\varsigma \over
dz}\right)^2 + \left({d \ln P \over dz}\right)^2 (1 + 2 \nabla^2),
\end{equation}
where $z_{\rm norm} =(\Re T_{\rm c}/\Omega_{\rm K}^2)^{1/2}$ is the vertical
scale height of the disc.

Because we have to solve the vertical structure for any given $T_{\rm c}$ and
$\Sigma$, $\alpha_{\rm eff}$ is not known {\it a priori}, but is a function
of $T_{\rm c}$ and $\Sigma$. In other words, there is one more boundary
condition than the number of differential equations, but one parameter is
unknown. This can be formulated in a more classical problem by
introducing a new equation:
\begin{equation}
{d\alpha_{\rm eff} \over dz} = 0.
\label{eq:alfconst}
\end{equation}
We solve a set of 7 equations (\ref{eq:stra}--\ref{eq:strc},
\ref{eq:griddef}, \ref{eq:phiconst}, \ref{eq:alfconst}) with 7 boundary
conditions in the optically thick approximation, while Eq.~(\ref{eq:deftau})
and the corresponding boundary condition are added to the problem in the grey
atmosphere approximation. We take all weights $\beta_i$ in
Eq.~(\ref{eq:discr}) equal to 0.5.

\subsection{Results}

\begin{figure}
\psfig{figure=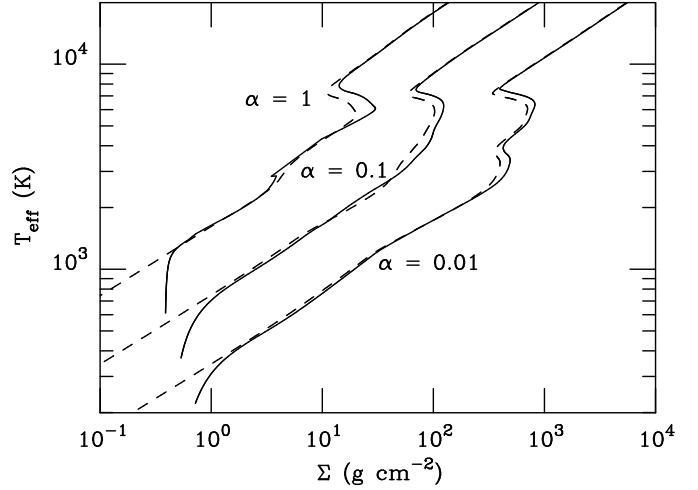,width=0.5\textwidth}
\caption{$\Sigma$ -- $T_{\rm eff}$ curves for $r = 10^{10}$ cm, $M_1$ = 1
M$\odot$, and $\alpha$ = 0.01, 0.1 and 1, in the grey atmosphere (solid
line) and optically thick (dashed line) approximations.}
\end{figure}

The equation of state of matter is interpolated from the tables of Fontaine
et al. \shortcite{fgv77}; in the low temperature regime (below 2000 K),
which is not covered by these tables, Saha equations are solved iteratively,
as described by Paczy\'nski \shortcite{p69}. The Rosseland mean opacities are
taken from Cox \& Tabor \shortcite{ct76} above 10,000 K, and from Alexander
\shortcite{a75} below. The Planck mean opacities are taken from Hur\'e
\shortcite{hu94}, and cover the range 1000 -- 30,000 K. The chemical
composition is assumed to be solar. We note that Liu \& Meyer-Hofmeister
\shortcite{liumey97} have shown that the use of improved (OPAL) opacities
does not much affect the results of the vertical structure calculations,
since most of the uncertainty resides in the $\alpha$-prescription.

Figure 1 shows an example of the $\Sigma$ -- $T_{\rm eff}$ ``S-curves'' we
obtained, in both the optically thick and the grey atmosphere approximations.
Differences are observed between the two cases, essentially when the disc is
very optically thin (i.e. $\Sigma$ less than about 1 g cm$^{-2}$), and when
the disc is convective. The values $\Sigma_{\rm max}$ and $\Sigma_{\rm min}$
which define the upper and lower stable branches can be fitted, in the grey
atmosphere approximation, by:
\begin{equation}
\Sigma_{\rm max} = 13.1 ~ \alpha^{-0.85} \left( {M_1 \over \rm M_\odot}
\right)^{-0.37} \left( {r \over 10^{10} \; \rm cm} \right)^{1.11}
~\rm g~cm^{-2}
\end{equation}
and
\begin{equation}
\Sigma_{\rm min} = 10.4 ~ \alpha^{-0.74} \left( {M_1 \over \rm M_\odot}
\right)^{-0.37} \left( {r \over 10^{10} \; \rm cm} \right)^{1.11}
~\rm g~cm^{-2}
\end{equation}
with the corresponding mass transfer rates:
\begin{equation}
\dot{M}_{\rm crit}^- = 4.0 ~ 10^{15} ~ \alpha^{-0.04} \left( {M_1 \over
\rm M_\odot} \right)^{-0.89} \left( {r \over 10^{10} \; \rm cm} \right)^{2.67}
~\rm g~s^{-1}
\end{equation}
and
\begin{equation}
\dot{M}_{\rm crit}^+ = 8.0 ~ 10^{15} ~ \alpha^{0.03} \left( {M_1 \over
\rm M_\odot} \right)^{-0.89} \left( {r \over 10^{10} \; \rm cm} \right)^{2.67}
~\rm g~s^{-1}.
\end{equation}
where $\dot{M}_{\rm crit}^-$ corresponds to $\Sigma_{\rm max}$ and 
$\dot{M}_{\rm crit}^+$ to $\Sigma_{\rm min}$.
We obtain in the optically thick case:
\begin{eqnarray}
\lefteqn{\Sigma_{\rm max} = 13.4 ~ \alpha^{-0.83} \left( {M_1 \over \rm
M_\odot} \right)^{-0.38} \left( {r \over 10^{10} \; \rm cm} \right)^{1.14}
~\rm g~cm^{-2}} \\
\lefteqn{\Sigma_{\rm min} = 8.3 ~ \alpha^{-0.77} \left( {M_1 \over \rm
M_\odot} \right)^{-0.37} \left( {r \over 10^{10} \; \rm cm} \right)^{1.11}
~\rm g~cm^{-2}} \\
\lefteqn{\dot{M}_{\rm crit}^- = 4.0 ~ 10^{15} ~ \alpha^{-0.004} \left( {M_1 \over
\rm M_\odot} \right)^{-0.88} \left( {r \over 10^{10} \; \rm cm} \right)^{2.65}
~\rm g~s^{-1}} \\
\lefteqn{\dot{M}_{\rm crit}^+ = 9.5 ~ 10^{15} ~ \alpha^{0.01} \left( {M_1 \over
\rm M_\odot} \right)^{-0.89} \left( {r \over 10^{10} \; \rm cm} \right)^{2.68}
~\rm g~s^{-1}}
\end{eqnarray}
These are in good agreement with values obtained previously (see e.g. Ludwig,
et al. \shortcite{lmr94}). A detailed comparison of
our $\Sigma-T_{\rm eff}$ curves with those obtained using the full radiative
transfer equation is left for a future paper \cite{ihls98}.

\section{Evolution of Dwarf Nova discs}

It is well known that models with constant values of $\alpha$ cannot
reproduce the observed light curves of dwarf novae \cite{sma84b}
(see however Faulkner et al., 1983).
They predict that globally, the disc reaches neither the hot nor
the cold branches. Instead a transition front propagates back and forth in a
relatively narrow zone, producing rapid and small amplitude oscillations of
the optical magnitude and the accretion rate. One can get around this
difficulty if one assumes that $\alpha$ is different on the hot and cold
branch of the S-curves. In the following, we assume $\alpha = \alpha_{\rm
hot}$ on the upper branch and $\alpha = \alpha_{\rm cold}$ on the lower
branch, with the following temperature dependence for $\alpha$:
\begin{eqnarray}
\log (\alpha)=\log(\alpha_{\rm cold})& +& \left[ \log(\alpha_{\rm hot})-
\log( \alpha_{\rm cold} ) \right] \nonumber \\ & &
\times \left[1+ \left( \frac{2.5 \times 10^4 \; \rm K}{T_{\rm c}} \right)^8
\right]^{-1} .
\label{eq:alpha}
\end{eqnarray}
The sharp transition between $\alpha_{\rm cold}$ and $\alpha_{\rm hot}$ is
required if one wishes to keep the values of $\Sigma_{\rm min}(\alpha)$ and
$\Sigma_{\rm max}(\alpha)$ in the effective S-curve equal to $\Sigma_{\rm
min}(\alpha_{\rm hot})$ and $\Sigma_{\rm max}(\alpha_{\rm cold})$.

\begin{figure}
\psfig{figure=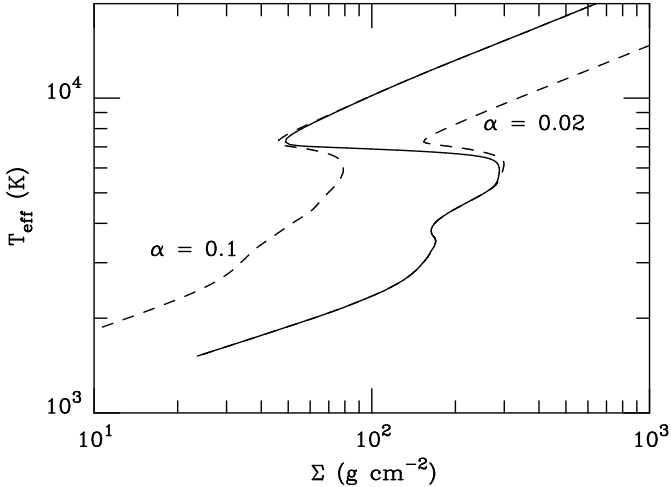,width=0.5\textwidth}
\caption{Modification of the $\Sigma - T_{\rm eff}$ curves when $\alpha$ is
no longer a constant. The dashed curves correspond to constant $\alpha$,
$\alpha = \alpha_{\rm hot}$ (left) and $\alpha = \alpha_{\rm cold}$ (right). The
solid line is obtained for $\alpha$ given by (\ref{eq:alpha}). The radius
of interest is 10$^{10}$ cm, and the primary mass 1.2 M$_\odot$}
\label{fig:scurves2}
\end{figure}

Figure~\ref{fig:scurves2} shows the changes introduced in the $\Sigma -
T_{\rm eff}$ curves when $\alpha$ is given by Eq.~(\ref{eq:alpha}).

Other $\alpha$ prescriptions, that we do not consider here, have been
proposed in the literature \cite{mm83,ccl95,vw96}, such as
$\alpha=\alpha_0(H/r)^{n}$ (see Lasota \& Hameury \shortcite{lh98} for a
discussion of this prescription).

In what follows we use $\alpha_{\rm cold}\sim 0.01$ as it is commonly
used in the literature. This assumption may be in contradiction with models
of turbulent viscosity in accretion discs \cite{gm97}.

In this article we only consider
the case of dwarf novae in which the disc extends down to the white dwarf
surface; its radius is taken from the Nauenberg \shortcite{n72}
mass-radius relation for white dwarfs.
The visual (V) magnitudes are computed assuming that each annulus of the 
disk emits blackbody radiation at the corresponding effective temperature.

\begin{figure}
\psfig{figure=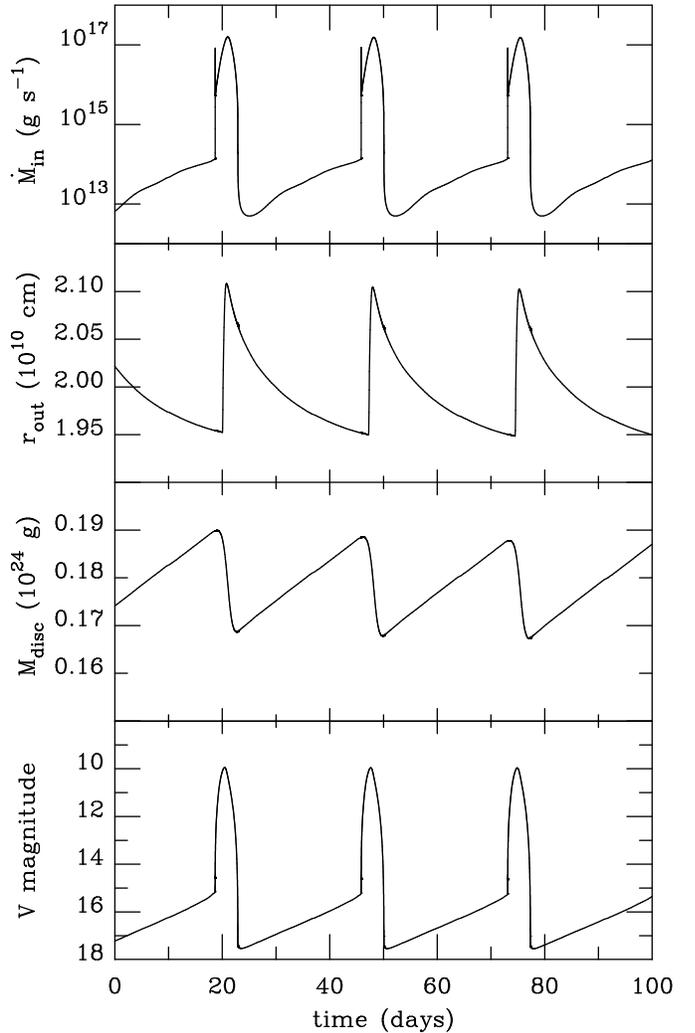,width=0.5\textwidth}
\caption{Outbursts properties for $M_1 = 0.6$ M$_\odot$, $r_{\rm in} = 8.5
\times 10^8$ cm, $\alpha_{\rm cold} = 0.04$, $\alpha_{\rm hot} = 0.20$,
$<r_{\rm out}> = 2 \times 10^{10}$ cm, and $\dot{M} = 10^{16}$ g s$^{-1}$. The
upper panel shows the mass accretion rate onto the white dwarf, the second
one the outer disc radius, the third one the disc mass, and the lower panel
the visual magnitude.}
\label{fig:ref_1}
\end{figure}

\begin{figure}
\psfig{figure=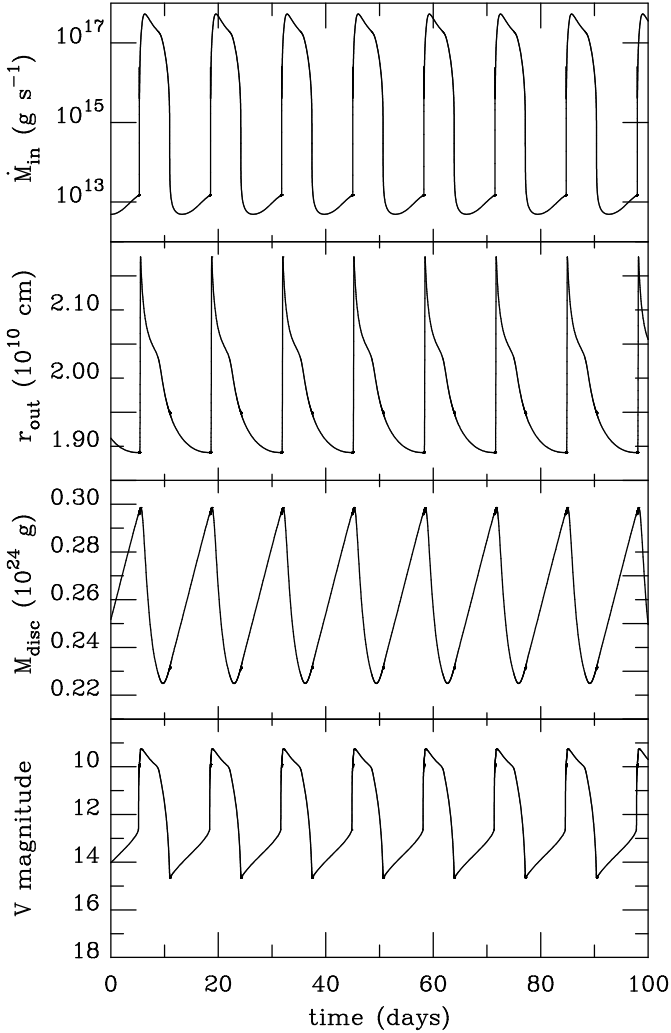,width=0.5\textwidth}
\caption{Same as fig. \ref{fig:ref_1}, but for $\dot{M} = 10^{17}$ g
s$^{-1}$.}
\label{fig:ref_10}
\end{figure}

Figures~\ref{fig:ref_1} and \ref{fig:ref_10} show two models obtained for
$M_1 = 0.6$ M$_\odot$, $r_{\rm in} = 8.5 \times 10^8$ cm, $\alpha_{\rm cold}
= 0.04$, $\alpha_{\rm hot} = 0.20$, an average $r_{\rm out}$ of $2 \times
10^{10}$ cm, and for mass transfer rates of $10^{16}$ g s$^{-1}$
(Fig.~\ref{fig:ref_1}) and $10^{17}$ g s$^{-1}$ (Fig.~\ref{fig:ref_10}). In
the first case, the disc experiences inside-out outbursts, while in the
second case, it experiences outside-in outbursts. The results concerning the
evolution of the disc outer radius are in good agreement, at least from a
qualitative point of view, with those previously obtained Ichikawa \& Osaki
\shortcite{io92}.

\begin{figure}
\psfig{figure=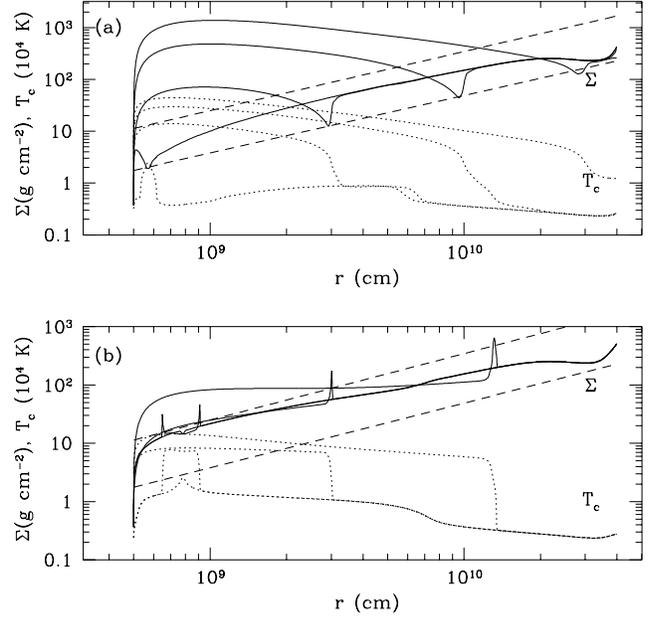,width=0.5\textwidth}
\caption{Typical profiles of the surface density $\Sigma$ (solid lines)
and the central temperature $T_{\rm c}$ (dotted lines) observed during the
evolution of the thin disc.
(a) Inward propagation of a cooling front and the associated density
rarefaction wave.
(b) Outward propagation of an inside-out heating front and the associated
density spike.
The dashed lines represent $\Sigma_{\rm min}$ (upper curve) and $\Sigma_{max}$
(lower curve).}
\label{fig:profhc}
\end{figure}

The typical surface density and central temperature radial profiles that we
observe in our simulations are shown in Fig.~\ref{fig:profhc}. During the
inward propagation of the cooling fronts, we recognize the density
rarefaction wave that precedes the region of rapid cooling, while during the
propagation of the heating fronts, we observe the density spike associated
with the sharp region of rapid heating \cite{pfl83,lpf85,can93a}. We also note
that our light curves are periodic (except when the number of points is so
small that numerical errors are large).

\subsection{Effect of the outer boundary condition}

We now take the same basic parameters as those used by Cannizzo
\shortcite{can93a} in his attempt to reproduce the light curve of SS
Cyg. We take $\alpha_{\rm hot}=0.1$, $\alpha_{\rm cold}=0.02$; the mass
transfer from the secondary is $\dot{M}_{\rm T}$ =10$^{-9}$ $M_{\odot} \ {\rm
yr^{-1}}$, and the mass of the accreting white dwarf is $M_1$ = 1.2 M$_\odot$
with an inner disc radius $r_{\rm in}=R_1=5 \times 10^8$ cm. We use the
optically thick approximation for determining the cooling function $Q^-$ of
the disc. Our reference model uses 800 grid points.

It is worth noting that, because we do not use exactly the same cooling
function (our values of $\Sigma_{\rm min}$ and $\Sigma_{\rm max}$ differ by
approximately 10 percent, and we do not use the same interpolation procedure
between the cold and the hot branches of the S-curves, where $\alpha$ varies
from $\alpha_{\rm cold}$ to $\alpha_{\rm hot}$), we do not expect to obtain
quite the same result as Cannizzo \shortcite{can93a}; in particular, our
model parameters have not been adjusted to fit SS Cyg's observed light curves.

\begin{figure}
\psfig{figure=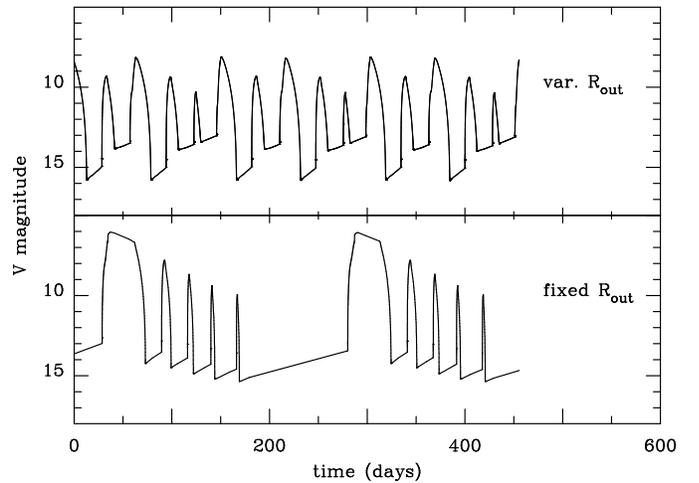,width=0.5\textwidth}
\caption{Calculated light curves assuming that the outer disc radius is fixed
(lower panel), or that it may vary with time (upper panel). The
average value of the outer disc radius is the same as in the fixed radius
case.}
\label{fig:comp_bc}
\end{figure}

Figure~\ref{fig:comp_bc} shows the two light curves obtained when the outer
disc radius is fixed at $r_{\rm out}=4 \times 10^{10}$ cm, and when $r_{\rm
out}$ is allowed to vary, the constant $c$ in Eq.~(\ref{eq:defT}) being
such that the average value of $r_{\rm out}$ is also $4 \times 10^{10}$
cm.

In the first case, we do reproduce the results of Cannizzo
\shortcite{can93a} qualitatively: we have an alternative
sequence of short and long
outbursts, which are all of the inside-out type. There are some quantitative
differences, which can be attributed to the differences mentioned above, but
the overall aspect is similar. In the second case, we obtain totally
different results which, not surprisingly, are similar to those of Ichikawa
\& Osaki \shortcite{io92}. We still obtain an alternating sequence of long
and short outbursts, but both the number of short outbursts and the cycle
length are quite different. As a rule, it is much easier to obtain such a
sequence of alternating short and long outbursts when the outer radius is
fixed than when it is allowed to vary.

When the outer edge of the disc is kept fixed at a constant value, we always
observe inside-out outbursts, as did Cannizzo \shortcite{can93a}, Ludwig et
al. \shortcite{lmr94} and Ludwig \& Meyer \shortcite{lm98}. This is in
contradiction with observations which show that, in a number of cases, the
instability is triggered in the outer parts of the disc. Such a discrepancy
has been attributed by Ludwig et al. \shortcite{lmr94} to the fact that the
disc might not extend to the white dwarf surface (note that this by itself
would not suffice to give outside-in outbursts, since the instability would
still be triggered in the innermost parts of the disc, but it could account
for the UV delay). They also argued that the viscosity law could be more
complex that the simple bimodal $\alpha$ prescription. However, when $r_{\rm
out}$ is allowed to vary, we obtain outside-in outbursts for mass transfer
rates high enough for the accumulation time of matter in the outer disc
regions to be shorter than the viscous diffusion time.

The qualitative difference observed between the two cases is due to
the fact that a much larger fraction of the disc is accreted during (large)
outbursts when one assumes that $r_{\rm out}$ is constant: the density
spike associated with the heating front brings some material to the outer
edge of the disc, which splashes on the rigid wall implied by the
boundary condition, and
thus strongly increases $\Sigma$; this matter will eventually be accreted
until the cooling wave starts, when $\Sigma = \Sigma_{\rm min}$ at
the outer edge of the disc; the accompanying rarefaction wave actually
brings $\Sigma$
below that value. The real situation is however quite different: when the
density spike reaches the disc outer edge, $r_{\rm out}$ increases (matter
carrying a large quantity of angular momentum flows outward), and the spike
finally dies out. Thus the cooling wave starts earlier, at a stage where less
mass from the outer disc has been accreted; moreover, during the quiescent
phase, the disc contracts (the outward angular momentum flux is reduced),
while mass transfer from the secondary continues. Both effects contribute to
increase $\Sigma$ in the outer disc and help the occurrence of outside-in
outbursts.

\subsection{Resolution Effects and the Disc Global Properties}

\begin{figure}
\psfig{figure=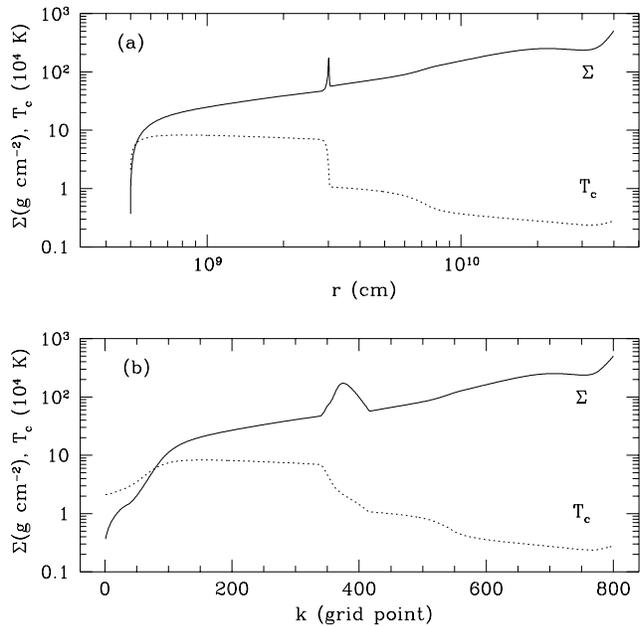,width=0.5\textwidth}
\caption{An example of the radial profiles of surface density $\Sigma$
(solid line) and central temperature $T_{\rm c}$ (dotted line) in the real
space (a) and the grid space (b),
using the adaptive grid at a resolution $N=800$
(reference model).
The strong gradients are resolved by the grid and about 100 grid points
are devoted to the resolution of the front itself.}
\label{fig:profkr}
\end{figure}

The use of adaptive grids is
an important improvement on previous time-dependent disc models in that
the transition fronts are always resolved,
whatever their location in the disc. This is illustrated in
Fig.~\ref{fig:profkr} where we show the sharpness of a heating front and
contrast it with the actual width ``seen'' by the adaptive grid. A fair
fraction of the grid ($\sim$ 100 of the total 800 grid points in our
reference model - adp800) is devoted to the front as a region of strong
gradients. Note that the inner edge of the disc is also resolved by the grid.

We now evaluate the effect that degrading the numerical resolution has on the
predicted light curves and on the variations of the total disc mass. We
compare the relative merits of our code in an adaptive grid version and
various fixed grid versions that we obtain when the grid function $W$ in
Eq.~(\ref{eq:griddef}) is defined as a function of $r$ only. In all cases,
the outer radius has not been allowed to vary. We tested r, log
and sqr grids (defined by a linear, a logarithmic and a square root spacing
between grid points, respectively), but we concentrate here only on the
results in the sqr grid case because this type of grid has been used
extensively in the literature. Indeed, the viscous equation has a
particularly simple form when the variables $x=r^{1/2}$ and $S = x \Sigma$
are used \cite{bp81}. We report in Table~\ref{tab:prevgrid} a few examples of
grids and resolutions that have been used in time-dependent studies so far.

\begin{table*}
\caption{A sample of previous time-dependent studies of thin accretion discs,
with the type of fixed grid and the numerical resolutions used. `System' refers
to a White Dwarf primary (WD, for the dwarf nova studies) and a black hole or
neutron star primary (BH, for the black hole or neutron star Soft X-ray
Transient studies). Note that Ichikawa \& Osaki (1992) and 
Smak (1994b) are the only studies in which the outer disc radius 
is allowed to move.}
\begin{center}
\begin{tabular}{cccc}
\hline Study & System & Type of grid & Num. Resolution \\ &&&\\
Smak \shortcite{sma84b} & WD & sqr & 20-25 \\
Lin, Papaloizou \& Faulkner \shortcite{lpf85} & WD &r/log & 35 \\
Mineshige \shortcite{min87} & WD & log & 401 \\
Mineshige \& Wheeler \shortcite{mw89} & BH & log & 21-41 \\
Ichikawa \& Osaki \shortcite{io92} & WD & r & 35-45\\
Cannizzo \shortcite{can93a} & WD &sqr & 25-200 \\
Cannizzo \shortcite{can94} & BH/WD & sqr & 300 \\
Cannizzo, Chen \& Livio \shortcite{ccl95} & BH & sqr & 103-1000 \\
Cannizzo \shortcite{can96} & WD & sqr & 400 \\
Cannizzo \shortcite{can98} & BH & r/sqr/log & 21-1000 \\
Ludwig \& Meyer \shortcite{lm98} & WD & ? & 200(500)\\
\\
\end{tabular}
\label{tab:prevgrid}
\end{center}
\end{table*}

\begin{figure}
\psfig{figure=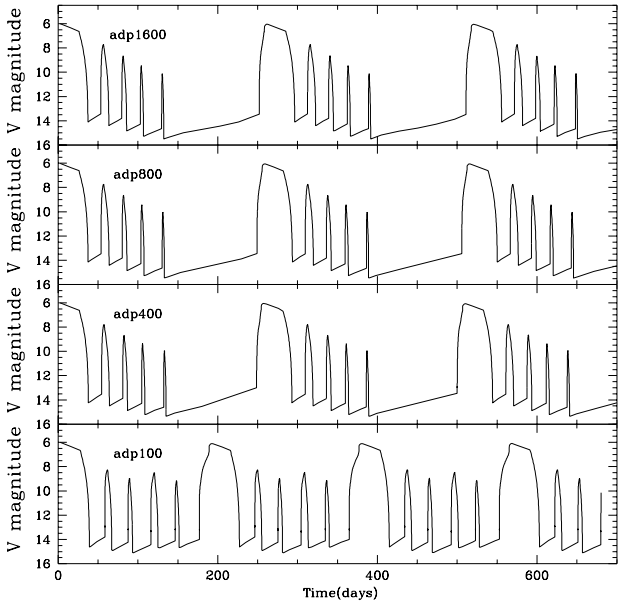,width=0.5\textwidth}
\caption{V magnitude light curves produced by our numerical model, using the
adaptive grid at various resolutions: adp100 (N=100), adp400 (N=400), adp800
(N=800), adp1600 (N=1600). The outer radius is fixed at $r_{\rm out} = 4
\times 10^{10}$ cm.}
\label{fig:mvadp}
\end{figure}

\begin{figure}
\psfig{figure=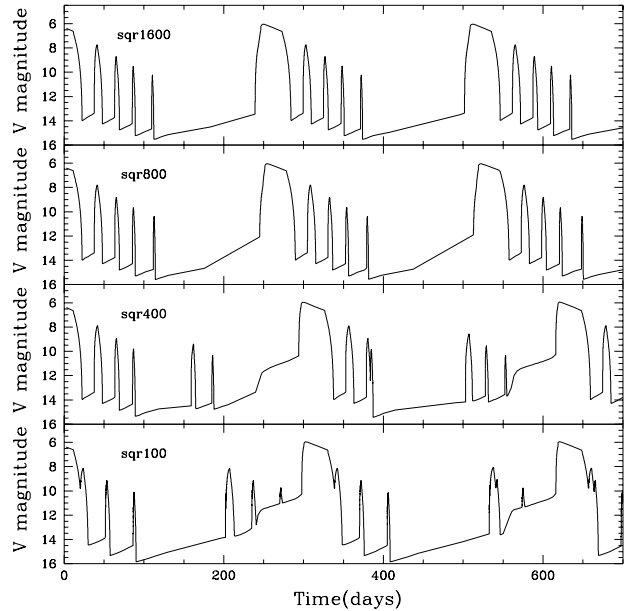,width=0.5\textwidth} \caption{V magnitude light
curves produced by our numerical model, using the fixed sqr grid at various
resolutions: sqr100 (N=100), sqr400 (N=400), sqr800 (N=800), sqr1600
(N=1600). $r_{\rm out}$ is not allowed to vary.}
\label{fig:mvsqr}
\end{figure}

\begin{figure}
\psfig{figure=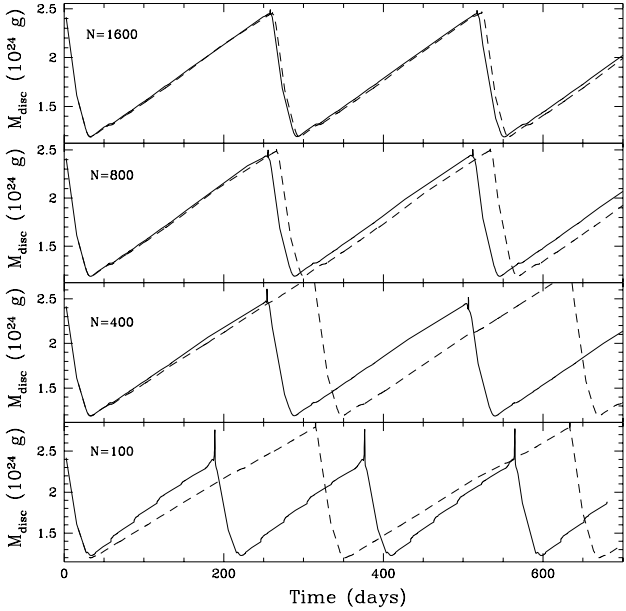,width=0.5\textwidth}
\caption{Evolution of the total disc mass with time for the various
test models at numerical resolutions $N=100,400,800,1600$. The adaptive
grid results are shown (solid lines) together with the fixed sqr grid
results (dashed lines). In all cases, $r_{\rm out}$ is fixed.}
\label{fig:mdiskcomb}
\end{figure}

We investigate the fixed sqr and adaptive grids at resolutions $N=100, 400,
800, 1600$ and we refer to these models as adp100 to adp1600 and sqr100 to
sqr1600. Figures~\ref{fig:mvadp} to \ref{fig:mdiskcomb} show the time
evolution of the V magnitude and the total disc mass in these models. A
straightforward conclusion is that a high numerical resolution ($N \geq 400$)
is required to reach a regime where the general shape of the outbursts
becomes independent of the resolution. This is particularly true for the
small outbursts, because the general aspect of the large ones is obtained at
a relatively moderate resolution. The adaptive grid naturally requires a
smaller number of grid points than the fixed grid to reach this regime
(typically 200 -- 400 points, compared with 800 points for the fixed
grid). It is worth noting that most results in the literature have been
obtained with resolutions which do not seem sufficient to avoid the
resolution limits (see Table~\ref{tab:prevgrid}). The inadequacy of such
a treatment is partially hidden by the fact that (1) the viscosity is a free
parameter which can be determined only by confronting the
time-dependent disc models with the observations, and (2) the cooling term
$Q^-$ that enters the thermal equation is not well known, and varies somewhat
from one author to another. These two facts may account for two differences
that we have with Cannizzo's \shortcite{can93a} results, even when we
take his assumptions, parameters and number of grid points. Our small
outbursts have peak amplitudes significantly smaller than the large outburst
peak amplitudes. We also find that the evolution of the disc mass is almost
independent on the resolution of the fixed sqr grid, in the sense that
$\sim 50$ \% of that mass is invariably accreted during large
outbursts, whatever the resolution.

\subsection{Resolution of the Transition Fronts}

Until recently, numerical simulations were not able to resolve the structure
of the transition fronts. For that reason, it has been very difficult to
compare the predictions of semi-analytical models with the numerical results,
and the physics of the fronts is not yet understood in much detail
\cite{lpf85,can94,can96,ccl95,vw96,v97,m84,m86}. We leave this comparison
for a future paper, and here instead we determine
the numerical resolution required
to address this problem properly. A heating or cooling front which is not
numerically resolved will be artificially enlarged; the physical results
obtained in these conditions are very questionable. In the extreme case
where a gradient (like a front) is limited to one numerical cell (between two
grid points), the fluxes across the gradient are likely to be reduced
and more of a numerical than a physical nature. Once again, one should keep
in mind that uncertainties in the thermal equation (\ref{eq:heat}), and in
particular in the radial flux $J$, may well hide resolution effects. For
instance, Cannizzo \shortcite{can93a} shows that the light curves of dwarf
novae do depend on the precise form of the thermal equation.

\begin{figure*}
\psfig{figure=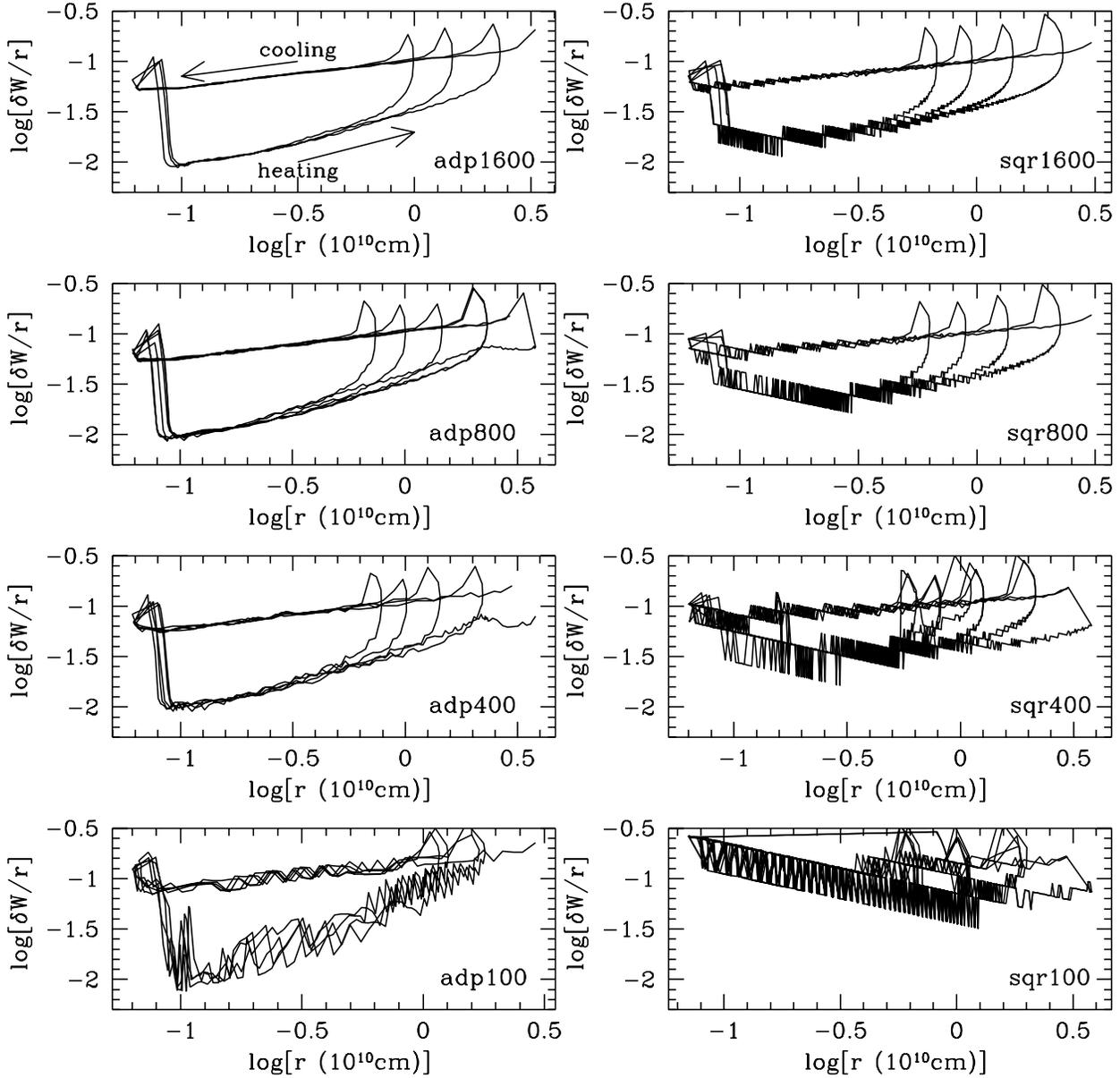,width=\textwidth}
\vskip -1truecm
\caption{A comparison of the evolution with time of the cooling and heating
front fractional widths ($\delta W/r$) as a function of the radius of
location of the fronts in the disc for several outbursts. The left panel
shows models with the adaptive grid at various resolutions, while the right
panel shows models with the fixed sqr grid at the same resolutions. The
inside-out heating fronts are sharper than the outside-in cooling fronts,
which means that the evolution of the disc is anti-clockwise in these
diagrams. The resolution effects are much more pronounced in the fixed sqr
grid case than in the adaptive grid case and even at high resolution
($N=1600$), the fixed sqr grid does not resolve heating fronts
better than marginally.
At lower resolution, the results on the front widths are
limited by resolution effects and are physically incorrect. At a resolution
of $N=100$, the front widths are clearly limited by the inter-grid spacing of
the fixed sqr grid.}
\label{fig:dw}
\end{figure*}

We show in Fig.~\ref{fig:dw} the fractional width, $\delta W/r$, of the
heating and cooling fronts during several outburst cycles of our test models.
Note that $\delta W$ is the width over which $90$~\% of the variation of
$\alpha$ (defined in Eq.~(\ref{eq:alpha})) occurs. This should not to be
taken as a definitive value for the front width, but rather as a typical
width (affected by resolution effects as well). What appears clearly in
Fig.~\ref{fig:dw} is that fixed sqr grids do not resolve transition fronts
well enough in the inner parts of the disc, even at high resolution. The
discontinuities seen in the sqr model widths are signatures of the resolution
limits and are very significant at low resolution ($N=100, 400$). The
situation is systematically worse for heating fronts, which are sharper than
cooling fronts. The adaptive grid has been devised to avoid large temperature
and density differences between two grid points, and predicts the correct
width for a moderate number of grid points $N$.

We also note that even when, close to the WD, the front widths are limited by
the resolution, the resulting light curves may still be correct
(sqr800-1600). The basic reason is that the general outburst shape is not
affected by what happens close to the compact object, where all
characteristic timescales are small. On the other hand, an adaptive grid may
approximately resolve the heating and cooling fronts, but produce incorrect
results at low resolution (for example when $N=100$). This is a result of
too many grid points being used in steep gradient regions, which leaves too
small a number of grid points in most parts of the disc.

\subsection{A Comparison of Numerical Grids}

We define the fractional inter-grid spacing (hereafter FIS) between two grid
points ($k$ and $k+1$) as:
\begin{equation}
\left( \frac{\Delta r}{r}
\right)_{k+1/2} \equiv \frac{r_{k+1}-r_k}{(r_{k+1}+r_k)/2},
\end{equation}
where the index $k$ refers to the grid point of interest in the grid extended
from $k=1$ at the inner edge of the disc to $k=N$ at the outer edge of the
discs.
A fixed $r$ grid is defined by:
\begin{equation}
r_{k+1}=r_k+\delta,
\label{eq:rgrid}
\end{equation}
$\delta$ being a constant depending on the resolution $N$ used. In this case,
the FIS is
\begin{equation}
\left( \frac{\Delta r}{r} \right)_{k+1/2} \simeq \frac{r_N-r_1}{N-1}\frac{1}{r_k}.
\end{equation}
A fixed sqr grid is defined by:
\begin{equation}
\sqrt{r_{k+1}}=\sqrt{r_k}+\delta
\label{eq:sqrgrid}
\end{equation}
and the corresponding FIS is
\begin{equation}
\left( \frac{\Delta r}{r} \right)_{k+1/2} \simeq \frac{2(\sqrt{r_N}-\sqrt{r_1})}
{N-1}\frac{1}{\sqrt{r_k}}.
\end{equation}
A log grid is defined by:
\begin{equation}
r_{k+1}=\beta r_k.
\label{eq:loggrid}
\end{equation}
The corresponding FIS is
\begin{eqnarray}
\left( \frac{\Delta r}{r} \right)_{k+1/2} & = & \frac{2(\beta-1)}{1+\beta} \\
\beta & = & \left( \frac{r_N}{r_1} \right)^{\frac{1}{N-1}}.
\end{eqnarray}

\begin{figure}
\psfig{figure=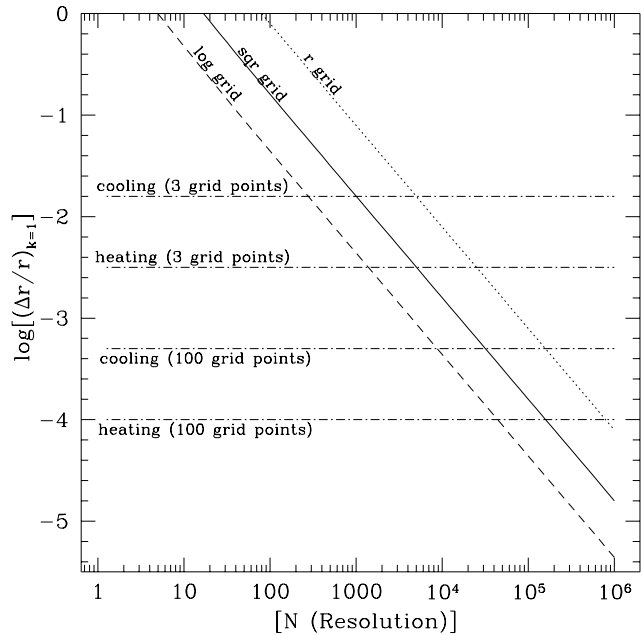,width=0.5\textwidth}
\caption{Dotted-dashed lines show the fractional inter-grid spacing ($\Delta
r/r$, also FIS) required by various fixed grids in order to resolve the
cooling and heating fronts of our reference model with $3$ and $100$ grid
points at the inner edge of the disc ($r_{\rm in}=5 \times 10^8$ cm), and
consequently everywhere in the disc (see text). The comparison with the
theoretical FIS of r (dotted line), sqr (solid line) and log (dashed line)
grids, as a function of the numerical resolution $N$, illustrate the very
high resolutions needed by fixed grids to resolve the transition fronts.}
\label{fig:gridresol}
\end{figure}

The adaptive grid runs show that the heating fronts reach a fractional width
as low as $\delta W/r \sim 10^{-2}$, and that $\delta W/r$ is $\sim
10^{-1.3}$ for the cooling fronts at the inner edge of the disc (see
Fig.~\ref{fig:dw}). This means that FISs of the order of ($10^{-1.8}$,
$10^{-3.3}$)/($10^{-2.5}$, $10^{-4}$) are required if a grid is
to resolve the cooling and heating fronts with $3$/$100$ grid
points respectively. We show in
Fig.~\ref{fig:gridresol} the FISs for the r, sqr and log grids as a function
of the numerical resolution $N$, and we compare them to the FISs required to
resolve the cooling and heating fronts with $3$ and $100$ points. The
comparison is performed at the inner edge of the disc ($k=1$) since fronts
resolved there will be resolved everywhere in the disc: the front fractional
widths increase with radius while the FISs of fixed grids are either constant
(log grid) or decrease with radius (r and sqr grids).

The fixed grids require very high numerical resolutions to resolve in detail
the fronts everywhere in the disc ($N \sim 10^4-10^6$). The situation is
worse when the accreting object is a black hole or a neutron star: the r and
sqr grids will resolve the fronts less and less as the inner radius of the
disc is reduced (see for example Cannizzo et al. \shortcite{ccl95}, Cannizzo
\shortcite{can98}). The log prescription has the advantage of defining a grid
with a FIS which does not vary with radius.

We conclude that reliable studies of the physical properties of transition
fronts cannot be completed in a reasonable
amount of computational time by fixed
grid codes. An exception (see Fig.~\ref{fig:dw}) is the structure of cooling
fronts, which can probably be addressed, at least in the outer parts of the
discs, by fixed sqr grid codes at high enough (but reasonable) numerical
resolutions.

\section{Conclusion}

We have constructed a numerical code which can calculate, in a reasonable
amount of computer time and at very high spatial resolutions,
long cycles of
accretion disc outbursts. This code works efficiently in the most general
framework of the disc instability model and does not require special
assumptions about viscosity or outer or inner radii. Its validity is of course
limited by the way physical processes such as turbulent viscosity,
convection, radiative transfer etc. are treated. Of course it is also a 1D
code modeling a fundamentally 2D (or even 3D) situation.

Since the mass of the central object enters the disc equations only in
$\Omega_K$, we expect most of our results to be valid for BH disc models as
well ({\i.e. similar}, but at a slightly smaller radius), as long as
irradiation and general relativistic effects can be neglected.

In future work we intend to include effects of irradiation (Dubus et al.
1998) and to apply the code to a systematic study of dwarf nova outbursts and
X-ray transient events.

\subsection*{Acknowledgments}
We are grateful to J. Cannizzo, R. Narayan, F. Meyer, B. Popham, J. Raymond
and J. Smak for useful discussions and to E. Quataert for comments on the 
manuscript. KM was supported by NASA grant NAG 5-2837.

\end{document}